\documentclass[letterpaper, 10 pt, conference]{ieeeconf}  

\usepackage{cite}
\usepackage{amsmath,amssymb,amsfonts}
\usepackage{algorithmic}
\usepackage{graphicx}
\usepackage{textcomp}
\usepackage{xcolor}
\usepackage{tabularx}
\usepackage{hyperref}
\hypersetup{
    colorlinks=true, 
    linkcolor=black,   
    citecolor=black,   
    urlcolor=black,    
}
\usepackage[utf8]{inputenc}
\usepackage[T1]{fontenc}
\usepackage{multirow}
\usepackage{pifont}
\usepackage{booktabs}
\usepackage{array}
\usepackage{makecell}

\makeatletter
\let\NAT@parse\undefined
\makeatother
\usepackage{hyperref}
\usepackage[numbers,sort&compress]{natbib}

\usepackage{amsfonts} 
\usepackage{amssymb} 
\usepackage{graphicx}
\usepackage{makecell}
\usepackage{amsmath}
\usepackage{CJK}
\usepackage{array} 
\usepackage{graphicx} 
\RequirePackage{CJKnumb}
\usepackage{caption}
\usepackage{graphicx, subfig}
\IEEEoverridecommandlockouts                              

\title{\LARGE \bf
Heterogeneous Space Fusion and Dual-Dimension Attention: A New Paradigm for Speech Enhancement*
}

\author{Tao Zheng$^{1}$, Liejun Wang$^{1\dagger}$ and Yinfeng Yu$^{1\dagger}$
\thanks{$^{\dagger}$Both Liejun Wang and Yinfeng Yu are corresponding authors.}
\thanks{$^{1}$Tao Zheng, Liejun Wang and Yinfeng Yu are with the School of Computer Science and Technology, Xinjiang University, Urumqi 830049, China (e-mail:{\tt\small 107552201375@stu.xju.edu.cn;
        wljxju@xju.edu.cn;
        yuyinfeng@xju.edu.cn;).}}%
\thanks{*This work was supported by these works: the Tianshan Excellence Program Project of Xinjiang Uygur Autonomous Region, China (2022TSYCLJ0036); the Central Government Guides Local Science and Technology Development Fund Projects (ZYYD2022C19); the National Natural Science Foundation of China under Grant 62303259;Graduate Student Research and Innovation Program in the Xinjiang Uygur Autonomous Region(XJ2024G089).}
}

\begin{document}

\bibliographystyle{unsrt}

\maketitle
\thispagestyle{empty}
\pagestyle{empty}

\begin{abstract}
Self-supervised learning has demonstrated impressive performance in speech tasks, yet there remains ample opportunity for advancement in the realm of speech enhancement research. In addressing speech tasks, confining the attention mechanism solely to the temporal dimension poses limitations in effectively focusing on critical speech features. Considering the aforementioned issues, our study introduces a novel speech enhancement framework, HFSDA, which skillfully integrates heterogeneous spatial features and incorporates a dual-dimension attention mechanism to significantly enhance speech clarity and quality in noisy environments.
By leveraging self-supervised learning embeddings in tandem with Short-Time Fourier Transform (STFT) spectrogram features, our model excels at capturing both high-level semantic information and detailed spectral data, enabling a more thorough analysis and refinement of speech signals. Furthermore, we employ the innovative Omni-dimensional Dynamic Convolution (ODConv) technology within the spectrogram input branch, enabling enhanced extraction and integration of crucial information across multiple dimensions. Additionally, we refine the Conformer model by enhancing its feature extraction capabilities not only in the temporal dimension but also across the spectral domain.
Extensive experiments on the VCTK-DEMAND dataset show that HFSDA is comparable to existing state-of-the-art models, confirming the validity of our approach.

\end{abstract}

\section{INTRODUCTION}

Speech communication is a fundamental mode of human interaction. However, in everyday speech communications, environmental noise, background noise, and other interfering factors frequently "pollute" the speech data, significantly diminishing its quality and clarity. Speech enhancement(SE) technology seeks to address this issue by minimizing the impact of noise while preserving the integrity of the original clear signal. Given its relevance across various practical scenarios, this paper primarily focuses on the study of single-channel speech enhancement.
\par

As technology advances, deep learning methods have gradually become the mainstream strategy in various fields of artificial intelligence. Numerous studies \cite{xu2014regression,li2019convolutional,li2020recursive,yu2021wea丨venet,jiao2024mfhca,guo2023rethinking},
have shown that deep learning models, with their excellent feature extraction and information modeling capabilities, exhibit significant potential across various domains. In the field of speech enhancement, the application of deep learning has led to notable advancements. Depending on the input method, speech enhancement techniques can be categorized into two types: one involves directly inputting the time-domain signal into the deep learning model, which facilitates rapid implementation but may require a complex network structure to effectively process the raw signal \cite{xiang2021nested,pandey2021dense}; the other method first processes the signal through Short-Time Fourier Transform (STFT), converting it into a time-frequency domain representation, before inputting it into the model. Within the time-frequency domain approaches, there are mainly two methods: mapping-based methods \cite{wang2020complex,li2021two} and masking-based methods \cite{hu2020dccrn,wang2022d}. 
In this paper, we focus on the study of masking-based SE methods.
\par
Self-supervised learning (SSL) models, due to their outstanding performance in various speech tasks, are considered a significant future direction for speech enhancement technology. However, the application of these models in the field of SE is still in its initial stages. Huang et al. \cite{huang2022investigating} have evaluated the performance of thirteen self-supervised models in SE tasks and proposed the strategy of directly applying self-supervised models to SE. Notably, when applied to SE tasks, SSL models often exhibit characteristics different from those in other speech-processing tasks. Huang and colleagues observed that detailed information might be lost in deeper layers of the network. Hung et al. \cite{hung2022boosting} confirm that using cascaded features as input significantly improves model performance in SE tasks.
\par
Attention mechanisms are more effective in handling sequential data and are, therefore, widely applied in speech tasks. Transformer utilizes multi-headed self-attention(MHSA) to process the sequence data \cite{vaswani2017attention}. While this model performs well in capturing global information, its effectiveness in extracting local information is limited. On the other hand, Convolutional Neural Networks(CNN) are proficient in modeling local data. The Conformer model \cite{gulati2020conformer} combines the strengths of both by integrating CNN and Transformers. 
 We observe that the one-dimensional convolutional module in the Conformer block not only extracts local information along the time dimension but also pays attention to relevant information along the frequency dimension, thereby enhancing the model's performance. We believe that the ability to focus on frequency-domain information also affects the model's performance, which may be one reason for this model's performance improvement.
\par
This study introduces a novel speech enhancement network that integrates heterogeneous spatial features (HSF) and incorporates a Dual-Dimension attention (DDA) mechanism. 
\par
The primary contributions of this paper are as follows: 

\begin{figure*}[t]
    \centering
    \includegraphics[width=1\textwidth]{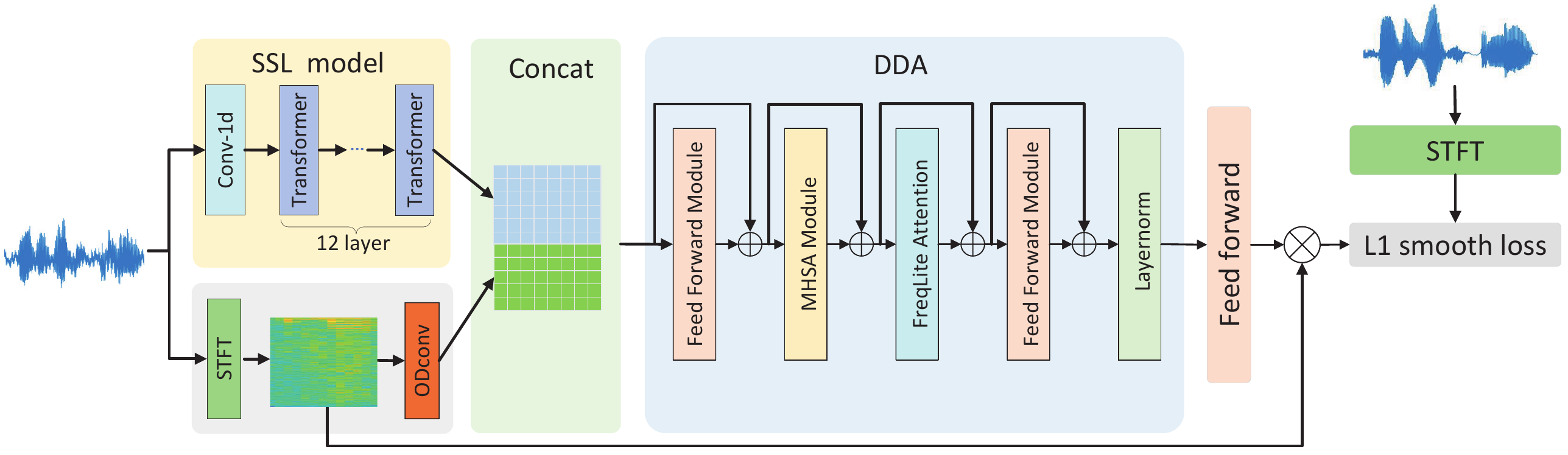}
    \caption{The architecture of our proposed method}
    \label{fig1}
\end{figure*}
\begin{itemize}

\item  Utilizes self-supervised embeddings combined with STFT spectrogram features to achieve the fusion of heterogeneous spatial features, allowing for simultaneous capture of high-level semantic and detailed information. 
\item  Implement Omni dimensional Dynamic Convolution (ODConv) \cite{li2022omni} technology in the spectrogram input branch to achieve integration and extraction of key information across all dimensions.
\item  Refine the conformer model to enhance its feature extraction capabilities not only in the temporal dimension but also in the spectral dimension. 

\end{itemize}

\section{related work}

\subsection{Self-Supervised Learning Models}

Self-supervised models have shown significant progress in speech tasks. The earliest methods, such as Contrastive Predictive Coding (CPC) \cite{oord2018representation}, and  Autoregressive Predictive Coding (APC) \cite{yang2022autoregressive}, first introduced unsupervised learning to audio pre-training. Building on these works, the wav2vec \cite{schneider2019wav2vec} series further enhanced the performance of automatic speech recognition (ASR). HuBERT \cite{hsu2021hubert} and WavLM \cite{chen2022wavlm} improved the performance and generalization ability of audio ASR. When applying pre-trained self-supervised models to downstream tasks, significant performance improvements have been observed through task-specific fine-tuning. For example, in the wav2vec 2.0 project, an SSL model, after being fine-tuned with labeled data using a CTC loss function, was used to enhance speech recognition tasks \cite{baevski2020wav2vec}. Research has shown that applying SSL models to speech-processing tasks extends beyond ASR. Specifically, studies in speech emotion recognition (SER) and SE have demonstrated improvements in model performance. For SER, Khare et al. \cite{khare2021self} utilized a transformer-based SSL model, enhancing performance by fine-tuning an initially trained transformer. In the field of SE, Lee et al. \cite{lee2024leveraging} explored strategically designed ensemble mapping processes within the SSL feature space, aiming to improve speech enhancement through adaptation strategies. Similarly, Song et al. \cite{song2023exploring} designed a regression-based variant of the WavLM objective, optimizing within an unsupervised learning framework to predict continuous outputs from masked regions of the input signals.

\subsection{Conformer}
When processing speech data, considering the context-related information of sequence data is crucial. Unlike traditional Recurrent Neural Networks (RNNs) or CNN, the Transformer model captures global context effectively by processing all positions in the input sequence simultaneously through its self-attention mechanism. Yu et al. employed a Transformer in the domain of speech enhancement and utilized LocalLSTM instead of positional embeddings to represent the local structure of speech signals \cite{yu2022setransformer}. Conformer enhances model performance by combining the advantages of CNN and Transformer. Kim et al. proposed a novel time-domain speech enhancement method named SE-Conformer \cite{kim2021se}. This method adopts the Conformer architecture and integrates it into the convolutional encoder-decoder (CED) framework to improve speech quality and clarity. 
Abdulatif et al. \cite{abdulatif2024cmgan} explore the efficacy of two-stage Conformer blocks in capturing temporal and spectral dependencies while maintaining a relatively low computational burden.

\section{method}

Fig. \ref{fig1} depicts the overall architecture of the model we propose. Speech data is fed using two branches of different spatial features. Specifically, one branch first transforms speech waveform data into a spectrogram via STFT and then uses ODConv technology to extract key information from the spectrogram. The other branch feeds the speech data into a self-supervised model to extract high-level semantic information. The two types of features are merged in the time dimension through a concatenation operation, which is then fed into a DDA module. This module is capable of extracting features across both the time and frequency dimensions. Finally, the data is processed through a Feedforward layer before output. The model employs an L1 smooth loss function to compute loss and optimize performance.
\begin{figure*}[t]
    \centering
    \includegraphics[width=1\textwidth]{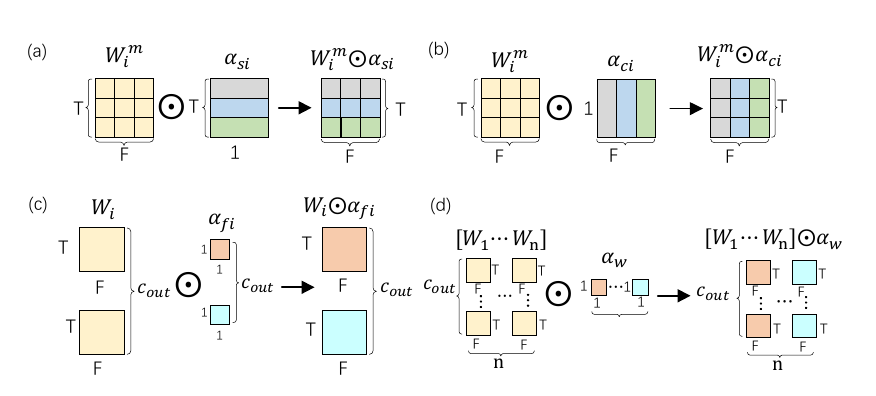}
    \caption{In ODConv, four distinct types of attention multiplication are progressively applied to convolutional kernels. Specifically, (a) denotes location-wise multiplication operations along the temporal dimension, (b) represents location-wise multiplication operations along the frequency dimension, (c) indicates channel-wise multiplication operations along the output channel dimension, and (d) corresponds to kernel-wise multiplication operations along the dimension of the convolutional kernel.}
    \label{fig2}
\end{figure*}

\begin{figure}[h]
    \centering
    \includegraphics[width=0.4\textwidth]
    {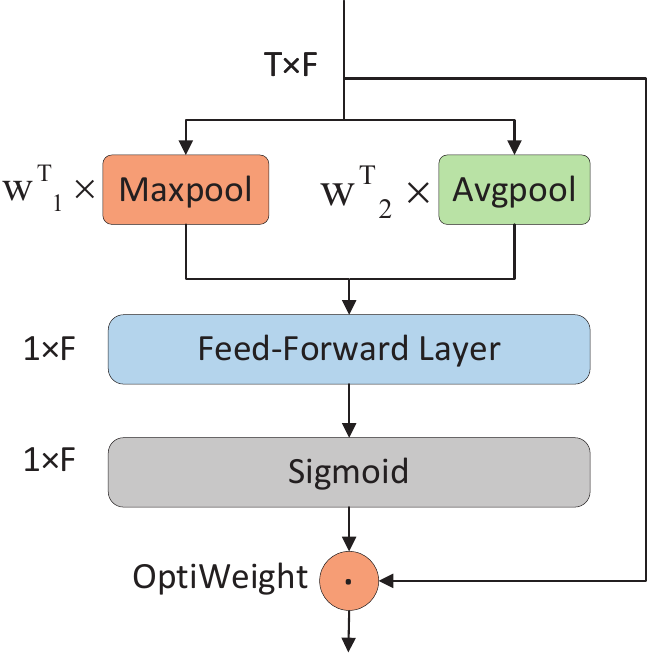}
    \caption{The FA module initially extracts features through two weighted pooling layers, producing 1×F dimensional attention weights. These weights are then fed into a feed-forward layer and processed through a sigmoid activation function for a nonlinear transformation. After activation, the attention weights are expanded to T×F dimensions, forming a relevance score. These attention scores are element-wise multiplied with the original T×F dimensional input to optimize the weights of various parts of the original input data. This output reflects the state of the input features after being adjusted by the attention mechanism.}
    \label{fig3}
\end{figure}

\subsection{ODConv Module} 
CNNs are frequently employed in SE tasks. Traditional CNNs use static convolution kernels, which cannot flexibly adapt to the diverse characteristics of input samples. Consequently, the exploration of dynamic convolution kernels represents a novel direction in CNN kernel research. Li et al. introduced the Omni-dimensional Dynamic Convolution (ODConv) \cite{li2022omni} and demonstrated its efficacy in object detection tasks. Motivated by this study, we have adopted ODConv for the processing of spectrograms. The ODConv mechanism, illustrated in Fig. \ref{fig2}, dynamically adjusts the weights of convolution kernels in response to the input data, thereby enhancing the model's ability to capture features effectively.

Specifically, ODConv is applied to STFT spectrograms after they undergo a time-frequency transformation. This is achieved by a parallel processing strategy that intricately tunes the convolution kernels across four dimensions. For a set of convolution kernels \(W_i\) in \(W = \{W_1, \ldots, W_n\}\),
we assign attention weights to the time dimension (\( \alpha_{si} \)), frequency dimension (\( \alpha_{ci} \)), output channels (\( \alpha_{fi} \)), and the overall convolution kernel (\( \alpha_{wi} \)) for each kernel \( W_i \).

The application of these attention weights occurs in a sequential manner, following the hierarchy of time, frequency, output channel, and convolution kernel scales. This sequential multiplication of the four attention types to the respective convolution kernel \( W_i \) is what grants the ODConv its distinctive flexibility and adaptability to multidimensional changes. Consequently, this enhances the network’s proficiency in feature extraction and information integration within complex scenarios.

The formal representation of this operation is given by the equation:
{\setlength\abovedisplayskip{0.4cm}
\setlength\belowdisplayskip{0.5cm}
\begin{equation}
\begin{split}
y=({\alpha }_{w1}\odot {\alpha }_{f1}\odot {\alpha }_{c1}\odot {\alpha }_{s1}\odot{W_1} +...\\+{\alpha }_{wn}\odot {\alpha }_{fn}\odot {\alpha }_{cn}\odot {\alpha }_{sn}\odot{W_2})*x,
\end{split}
\end{equation}}where \( \odot \) denotes element-wise multiplication, and \( * \) represents the convolution operation applied to the input \( x \). This equation encapsulates the core principle of ODConv—weighting convolution kernels by paying attention to different dimensions before applying them to the input, thereby adapting the process to the data's unique features.

\subsection{DDA MODULE}
The conformer architecture is a commonly used architecture for speech tasks, and we observe that the one-dimensional convolutional module in the conformer block can play a role in extracting localized information in the time dimension while also paying some attention to information related to the frequency dimension, thus improving the performance of the model, and hence it can be seen that the ability of the model to pay attention to information in the frequency dimension also affects the performance of the model. We believe that attention in the frequency dimension may be more competent for such a task. At the same time, based on the consideration of lightweight, we designed the FreqLite Attention (FA) module for feature extraction of frequency dimension speech information.

\par

Fig.\ref{fig3} depicts the FA module, which is a lightweight attention mechanism designed to focus on the frequency dimension. For the input features \( X \in \mathbb{R}^F \), the FA module re-computes a relevance score \( U_t \) for each element \( X_t \), Here is the formula for the attention fraction \( U_t \):
\begin{equation}
	\begin{split}
	&{U}_{t}= Attention({X}) \\
	&=\sigma [{\omega }_{1}^{T}AvgPool(X)+ {\omega }_{2}^{T}MaxPool(X)],
	\end{split}
\end{equation}\( \sigma \) acts as an activation function, while \( \mathbf{\omega}_{1} \) and \( \mathbf{\omega}_{2} \) represent the learned weight matrices. The operations AvgPool and MaxPool perform average pooling and max pooling on the input \( X \), respectively, aiding in capturing frequency information at various levels. Specifically, average pooling uniformly considers all frequency components, which helps in capturing the overall frequency distribution. Conversely, max pooling emphasizes the most prominent frequency component, thus focusing on key frequencies relevant to the task. The outputs from these two pooling strategies are integrated through a dynamic weighting system, allowing the model to adjust its focus on frequency features according to the specific requirements of the task. This adaptive structure not only enhances the model's sensitivity to frequency details but also improves computational efficiency, making the FA module a powerful tool for managing frequency complexity in various applications.

\begin{table}[h]
 \renewcommand{\arraystretch}{1.3}
\centering
\resizebox{\linewidth}{!}{
}

\caption{Comparison of experimental results}
\label{table1}
\scalebox{0.9}{
\begin{tabular}{|l|c|c|c|c|c|}

\hline
\textbf{Model}        & \textbf{PESQ $\uparrow$} & \textbf{CSIG $\uparrow$} & \textbf{CBAK$\uparrow$} & \textbf{COVL$\uparrow$} & \textbf{STOI$\uparrow$} \\ \hline
\textbf{HFSDA(ours)}                & \textbf{3.28} & \textbf{4.56} & \textbf{3.63} & \textbf{3.91} & \textbf{0.959}         \\\hline
BSSE \cite{hung2022boosting}                  & 3.20          & 4.53          & 3.60          & 3.88          & -             \\
SE-SSRA  \cite{khare2021self}             & 2.46          & 3.53          & 3.10          & 2.98          & -             \\\hline
T-GSA  \cite{kim2020t}               & 3.06          & 4.18          & 3.59          & 3.62          & -             \\
SE-T \cite{yu2022setransformer}                  & 2.62          & -             & -             & -             & 0.93          \\ 
SE-Conformer \cite{kim2021se} & 3.13  & 4.45          & 3.55          & 3.82          & 0.95          \\ \hline
MGAN-OKD \cite{shin2023metricgan}  & 3.24  & 4.53          & 3.65          & 3.91          & 0.950         \\ 
MANNER \cite{park2022manner} & 3.21  & 4.45          & 3.55          & 3.82          & 0.95          \\ 
DeepFilterNet3 \cite{schroter2023deepfilternet} & 3.17  & 4.34          & 3.61          & 3.77          & 0.94          \\ 
PFPL \cite{hsieh2020improving} & 3.15  & 4.18          & 3.60          & 3.67          & 0.950          \\ \hline
\end{tabular}
}
\end{table}

Subsequently, \( U_t \) is expanded to match the dimensions of \( X \), and then element-wise multiplied with the original input \( X_t \) to yield the adjusted \( X_t \).
\begin{equation}
 {X}_{t}= expand({U}_{t})*X \\,
\end{equation}the element-wise multiplication, denoted by *, involves multiplying the input features by the weight coefficients 
\( U_t \), effectively weighting the original input.

The FA module replaces the convolutional module within the conformer block. This newly devised attention model incorporates multi-head self-attention (MHSA) 
 along the temporal dimension and FreqLite Attention (FA) along the frequency dimension, enabling Dual-Dimension attention (DDA). This modification not only reduces the parameter count compared to the original conformer module but also enhances the model's performance by effectively capturing both temporal and frequency features.

\section{ EXPERIMENTAL SETUP}
\subsection{ Dataset and Assessment of indicators} 

Our study utilized the commonly employed VCTK-DEMAND dataset, which comprises a mix of mixed noise and clean speech, to evaluate the denoising performance of our model. The clean speech recordings were sourced from the VoiceBank corpus. The training set consists of 11,572 audio recordings, while the test set includes 872 recordings. The noise data were obtained from the DEMAND database, with the training dataset featuring ten different types of noise, such as babble, cafeteria, and kitchen noise, with Signal-to-Noise Ratios (SNRs) of 0, 5, 10, and 15 dB. The test dataset contained five types of noise, with SNRs of 2.5, 7.5, 12.5, and 17.5 dB. There is no overlap of noise data or noise conditions between the training and test sets. Our evaluation metrics include Perceptual Evaluation of Speech Quality (PESQ), CSIG, CBAK, COVL, and the Short-Time Objective Intelligibility (STOI) measure. 
\subsection{Experimental setup}

\begin{table}[ht]

\centering
\renewcommand{\arraystretch}{1.3}

\caption{Ablation experiments on individual modules}
\label{table2}
\scalebox{0.9}{
\begin{tabular}{|c|c|c|c|c|c|}

\hline
\textbf{Model}        & \textbf{PESQ $\uparrow$} & \textbf{CSIG $\uparrow$} & \textbf{CBAK$\uparrow$} & \textbf{COVL$\uparrow$} & \textbf{STOI$\uparrow$} \\ \hline
\textbf{\makecell[l]{WavLM\\+STFT(ODConv)\\+DDA}}& \textbf{3.28} &\textbf{ 4.56} & \textbf{3.63} & \textbf{3.91} & \textbf{0.959} \\ \hline
\makecell[l]{WavLM+\\+STFT(ODConv)\\+conformer(+FA)} & 3.24 & 4.52 & 3.60 & 3.90 & 0.957 \\ \hline
\makecell[l]{WavLM+\\+STFT(ODConv)\\+conformer}  & 3.22 & 4.53 & 3.62 & 3.89 & 0.955 \\ \hline
\makecell[l]{WavLM\\ + FAblock} & 3.12 & 4.31 & 3.47 & 3.78 & 0.945 \\ \hline
\makecell[l]{Wav2vec+\\STFT(ODConv)\\+DDA} & 3.14 & 4.48 & 3.56 & 3.83 &0.957 \\ \hline
\makecell[l]{STFT(ODConv)+ \\FAblock} & 3.10 & 4.35 & 3.40 & 3.62 & 0.948\\ \hline
\makecell[l]{STFT+ \\FAblock} & 2.98 & 4.22 & 3.51 & 3.53 & 0.941 \\ \hline
\end{tabular}

}

\end{table}

During the data preprocessing stage, speech from the training set was segmented into 1.5-second slices to facilitate model processing; meanwhile, speech in the test set was kept at its original length to evaluate the model's performance on data of varying lengths. For spectral feature extraction, we employed a 25-millisecond window length (corresponding to a 400-point Fast Fourier Transform (FFT) and a 10-millisecond step size using a Hamming window. Consequently, the resultant spectrograms have 200 frequency bins in the frequency dimension, while the length in the time dimension  depends on the duration of each audio track. The model utilized two DDA blocks ($N=2$). During training, the batch size was set to 16, and the Adam optimizer was used for parameter updates. The training comprised 200 epochs, with an initial learning rate set at $1 \times 10^{-4}$. Additionally, a learning rate scheduler was implemented, adjusting the learning rate by a decay factor of 0.5 every ten epochs.
\subsection{Performance Comparison}
In this study, we first compared our proposed model with the current advanced enhancement models that have applied self-supervised learning to the VCTK dataset. The results show that our model outperforms all reference baselines in terms of performance. Subsequently, we further compared our model with models that employ Transformer and Conformer architectures. The analysis indicates that our model also exhibits superior performance. Considering that most current models do not incorporate self-supervised learning or the Conformer architecture, we also compared our model with other models that use different network architectures. The comparative results also demonstrate the competitiveness and superiority of our model.

\par
Table \ref{table1} presents a comparative analysis of our newly developed HFSDA module against several established speech enhancement models, evaluating them across various performance metrics. Models such as BSSE~\cite{hung2022boosting}, SE-SSRA~\cite{lee2024leveraging}, and T-GSA~\cite{kim2020t} incorporate self-supervised learning techniques. In contrast, SE-T~\cite{yu2022setransformer} and SE-Conformer~\cite{kim2021se} leverage multi-head self-attention mechanisms. Additionally, the models MGAN-OKD~\cite{shin2023metricgan}, MANNER~\cite{park2022manner}, DeepFilterNet3~\cite{schroter2023deepfilternet}, and PFPL~\cite{hsieh2020improving} are recognized for their exceptional performance in the enhancement domain, as evidenced by recent research. This table facilitates an in-depth understanding of how our HFSDA module compares to existing technologies in terms of enhancing speech quality.

\par

\subsection{Ablation Analysis}
As shown in Table \ref{table2}, ablation studies were meticulously executed to substantiate the functional significance of the constituent modules in our architecture. The initial substitution of the DDA module with a conformer module led to a decrease in the PESQ score by 0.08, accompanied by declines in other evaluative metrics. These findings suggest the potential superior suitability of the lightweight FrquLite Attention module over traditional convolutional modules in speech enhancement tasks. Subsequently, the FA module was integrated between the MHSA and CNN components within the conformer architecture; however, this modification did not yield a significant improvement in model performance. This indicates that the frequency information, once weighted and reshaped by the FA module, may no longer be conducive for processing by the CNN. Concurrently, differential self-supervised models were employed in our experimentation; the substitution of WavLM with Wav2Vec led to a moderate decrement in model performance, thereby attesting to the comparative advantage of WavLM in the realm of speech enhancement. To assess the impact of heterogeneous space fusion features on model prowess, inputs were confined to discrete spatial features. Sole reliance on SSL output manifested a decrease of 0.24 in PESQ, whereas exclusive dependence on STFT output entailed a PESQ reduction of 0.18, corroborating the substantial influence of heterogeneous space fusion features on model competence. Lastly, the removal of the ODConv component from a singular branch model precipitated a PESQ decrease of 0.30, underscoring the indispensability of the initial processing of the STFT spectrogram.

\section{CONCLUSIONS}
In this paper, we introduce HFSDA, a novel speech enhancement model that addresses the complexities associated with noisy communication environments, utilizing heterogeneous spatial features and a dual-dimension attention mechanism. Our model effectively combines self-supervised embeddings with spectrogram features derived from STFT, enabling the subtle integration of semantic and detailed acoustic information. This holistic approach significantly enhances the clarity and quality of speech signals. Overall, our extensive evaluations of the VCTK-DEMAND dataset firmly demonstrate the effectiveness and superiority of our proposed method across several critical metrics, including PESQ and STOI. Spatial feature fusion in the field of speech enhancement has not yet seen significant development. Our research may pave the way for novel approaches to spatial feature fusion in the future.

\small
\bibliography{ref}
\end{document}